\documentclass[journal]{IEEEtran}

\usepackage{amsmath,amsfonts}
\usepackage{graphicx}
\usepackage{cite}
\usepackage{url}
\usepackage{xcolor}

\title{Beyond Algorithms: Conceptual Innovation in Medical Imaging AI}
\author{Mark A.~Anastasio~\IEEEmembership{Fellow,~IEEE}%
\thanks{Mark A. Anastasio is with the Mallinckrodt Institute of Radiology and Department of Electrical \& Systems Engineering, Washington University in St. Louis, St. Louis, MO 63110 USA (e-mail: anastasio@wustl.edu).}}

\begin{document}
\maketitle

\begin{abstract}
Artificial intelligence has driven rapid progress in medical imaging research, producing increasingly sophisticated algorithms and steady improvements on benchmark tasks. However, this algorithm-centric trajectory has also revealed a growing imbalance: while computational methods advance rapidly, the conceptual foundations that define imaging tasks, evaluation metrics, and clinical meaning sometimes remain underexamined. In this Perspective, we distinguish \emph{algorithmic innovation}, which focuses on improving computational implementations and performance within a fixed problem definition, from \emph{conceptual innovation}, which reframes what problems are posed, how success is measured, and why an approach is clinically relevant. We argue that prevailing incentive structures, training pathways, and publication norms disproportionately reward algorithmic novelty, particularly for early-career researchers, while at times undervaluing conceptual contributions that are essential for scientific maturation and clinical translation.
 Through representative examples from medical imaging AI, we show how insufficient conceptual grounding can lead to misaligned objectives, fragile generalization, and limited real-world impact. We conclude with actionable recommendations for researchers, mentors, reviewers, and journals to better recognize, support, and integrate conceptual innovation alongside algorithmic advances.
\end{abstract}


\section*{Contribution Statement}
This Perspective makes three primary contributions. It provides an operational distinction between \emph{conceptual innovation} and \emph{algorithmic innovation} in AI-based medical imaging, identifies practical indicators of conceptual innovation that can help authors, reviewers, and editors assess contributions beyond algorithmic novelty, and examines systemic factors that discourage conceptual work---particularly for early-career researchers---while offering recommendations for better aligning incentives with the rigor and clinical impact expected of medical imaging science.

\section{Introduction}
Over the past decade, artificial intelligence (AI) has transformed medical imaging research \cite{litjens2017survey,hosny2018radiology}. Advances in machine learning architectures, optimization strategies, and computational resources have enabled automated image reconstruction, segmentation, detection, and classification at unprecedented scales and levels of performance. Benchmark-driven progress has accelerated discovery and lowered technical barriers, positioning AI as a central driver of innovation across imaging modalities and future clinical applications \cite{litjens2017survey,hosny2018radiology,knoll2020fastmri}.


Despite these successes, a substantial portion of method-driven and engineering-focused medical imaging AI literature reports incremental algorithmic improvements on narrowly defined tasks \cite{lipton2019troubling,sculley2018winners,maierhein2024metrics}. By comparison, broader questions about which problems are being addressed, how success is defined, and whether these objectives align with meaningful clinical decisions often receive less explicit emphasis. This imbalance can lead to progress that is well quantified yet only weakly connected to scientific understanding or clinical impact.

Situating contemporary medical imaging AI within the longer tradition of imaging science provides a useful perspective for examining this gap, clarifying how advances in learning-based methods interact with established principles of problem formulation, evaluation, and interpretation. This motivates a clearer distinction between two complementary forms of innovation, algorithmic innovation and conceptual innovation, and a more explicit account of how each should be recognized. Our goal is not to re-litigate broader debates about enthusiasm or skepticism surrounding AI, but to provide an operational vocabulary and review-ready criteria for recognizing conceptual contributions alongside algorithmic advances.

\section{Conceptual and Algorithmic Innovation}

Conceptual innovation concerns \emph{what} problems are posed, \emph{how} they are framed, and \emph{why} they matter. It introduces new mental models, reframes objectives, and clarifies assumptions that shape entire lines of research. In medical imaging AI, conceptual choices determine task formulation, target variables, and evaluation criteria, thereby constraining what algorithmic optimization can meaningfully achieve. From this perspective, conceptual innovation functions as \emph{upstream scientific control}: decisions made at the conceptual level govern what constitutes an error, which trade-offs are visible, and which failure modes can be detected.

It should be noted that conceptual innovation can occur at multiple levels. In some cases, as described above, it involves changes in task framing, such as redefining the quantity to be estimated, the criteria for success, or the role of the output in downstream decision-making. In other cases, conceptual change occurs at the level of epistemology or methodology, for example by altering how prior knowledge is represented or how data are used in learning. While both forms are meaningful, this paper focuses primarily on conceptual innovation at the level of task framing, as these choices most directly determine scientific validity and clinical relevance.

\begin{figure}[t]
\centering
\includegraphics[width=3in]{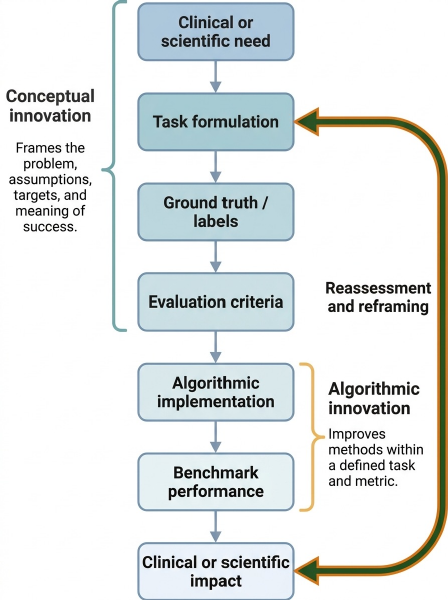}
\caption{Conceptual and algorithmic innovation in medical imaging AI. Conceptual innovation acts upstream by shaping the clinical or scientific question, target variable, assumptions about ground truth, and evaluation criteria. Algorithmic innovation acts downstream by improving computational mechanisms within a given problem formulation.}
\label{fig:workflow}
\end{figure}

Algorithmic innovation, by contrast, concerns how a defined problem is solved. It encompasses advances in architectures, loss functions, training strategies, and related computational mechanisms that improve performance within a fixed task definition and evaluation framework. While essential, algorithmic advances cannot compensate for misaligned problem definitions or inappropriate evaluation criteria. The two forms of innovation are therefore complementary but not interchangeable: conceptual innovation primarily shapes the scientific validity and clinical relevance of a task by influencing how objectives, assumptions, and evaluation criteria are defined, whereas algorithmic innovation primarily shapes how efficiently and accurately that task is executed within a given framing.
 This relationship is summarized schematically in Fig.~\ref{fig:workflow}, which emphasizes that conceptual innovation acts upstream by shaping task definitions, assumptions, and evaluation criteria, whereas algorithmic innovation operates within a defined problem formulation.

In practice, conceptual and algorithmic innovation are often intertwined. Algorithmic developments frequently serve as delivery mechanisms through which conceptual advances are instantiated, tested, and communicated within a quantitative research culture. In such cases, the algorithm may not constitute the primary intellectual contribution, but instead operationalizes a conceptual shift, such as redefining image quality in task-based terms, treating uncertainty as a first-class model output, or reframing imaging from diagnostic classification to prognostic or decision-support roles. Multiple algorithmic implementations may be capable of delivering the same conceptual insight, and the durability of the contribution often lies in the conceptual framing rather than in the specific computational mechanism employed, particularly when task definitions and evaluation criteria are central to validity.

Introducing a new learning procedure or network architecture does not, by itself, constitute conceptual innovation; such contributions remain algorithmic or implementation-level innovations when they improve performance under an existing task definition. Network design becomes conceptually innovative only when it operationalizes a change in problem framing, for example, by redefining the quantity being estimated, the criteria by which success is measured, or the interpretation of model outputs. Similarly, modifying an objective function constitutes algorithmic innovation when it improves model performance under an existing notion of success, but becomes conceptual innovation when it redefines what success means or how performance relates to scientific or clinical objectives. In this sense, algorithmic mechanisms may embody conceptual advances, but novelty in computational design alone is insufficient to establish a conceptual contribution. Because many studies span both categories, the distinction is intended to clarify the primary locus of contribution (problem framing versus computational mechanism) and to guide appropriate attribution and evaluation.

Conceptual innovation can be recognized through several recurring features. These include reframing the research question, making implicit assumptions explicit, introducing alternative evaluation criteria, exhibiting method-agnostic generality, enabling new lines of inquiry, and influencing research practice and discourse. Because these features are useful for authors, reviewers, and editors when assessing contributions beyond algorithmic novelty, they are summarized in Box~1.

It is also important to note that conceptual refinement is not always upstream of algorithmic progress. In many cases, algorithmic innovation has itself exposed conceptual gaps by amplifying limitations that were previously understood in principle but less visible in practice. In image reconstruction, for example, it has long been recognized that pixel-based fidelity metrics such as PSNR or SSIM are imperfect proxies for diagnostic utility \cite{barrett2004foundations,wang2009mse,wang2004ssim,haldar2026state}. However, the advent of highly expressive deep learning reconstruction methods has made this insufficiency more consequential and more apparent: models can achieve excellent scores under conventional metrics while producing visually plausible images that exhibit structured distortions, suppress diagnostically relevant features, or behave unstably under modest changes in sampling or anatomy \cite{antun2020instabilities,bhadra2021hallucinations}. Rather than revealing a new conceptual problem, these behaviors have sharpened and operationalized a long-standing one, prompting renewed scrutiny of assumptions about ground truth, perceptual similarity, and evaluation criteria. In this way, algorithmic advances can act as ``stress tests'' for prevailing task definitions, catalyzing conceptual refinement in response.

\section*{Box 1. Identifying Conceptual Contributions}
Conceptual contributions may be identified by asking whether a study:
\begin{itemize}
\item \textbf{Reframes the research question} by altering what quantity is estimated, which decisions are supported, or how imaging outputs are used in downstream clinical or scientific contexts.
\item \textbf{Makes implicit assumptions explicit} by interrogating sources of ground truth, label stability, data representativeness, or the relationship between proxy metrics and underlying clinical or scientific objectives.
\item \textbf{Introduces alternative evaluation criteria} that redefine success in terms of decision performance, uncertainty, robustness, or downstream impact rather than incremental improvements on established benchmarks alone.
\item \textbf{Exhibits method-agnostic generality}, such that the contribution applies across multiple algorithms, architectures, imaging modalities, or clinical settings rather than being tied to a specific implementation.
\item \textbf{Enables new lines of inquiry} by exposing previously unrecognized failure modes, generating new research questions, or establishing frameworks that subsequent studies can adopt and extend.
\item \textbf{Influences research practice and discourse} by shaping how problems are formulated, how results are interpreted, or how evaluation norms evolve within the field, even if immediate performance gains are modest or absent.
\end{itemize}

\section{Conceptual Innovation Across Medical Imaging Subfields}

Conceptual innovation has historically played a central role in many traditional subfields of medical imaging, including imaging physics, image reconstruction, task-based image quality assessment, and quantitative imaging \cite{barrett2004foundations,sullivan2015metrology,kessler2015emerging}. In these areas, progress has depended critically on explicit problem formulation, careful definition of quantities to be estimated, and principled evaluation strategies. Ill-posedness, measurement uncertainty, and the need for interpretability made underlying assumptions unavoidable, forcing researchers to engage directly with conceptual questions before algorithmic solutions could be meaningfully developed.

For example, advances in imaging physics and inverse problems were often driven by rethinking forward models and estimation targets, and, in some cases, by introducing new regularization assumptions that altered how prior knowledge and uncertainty were represented, rather than by incremental improvements to numerical solvers alone \cite{kak2001principles,engl1996regularization,arridge2019solving}. Similarly, task-based image quality assessment reframed image quality as a property defined by observer or decision performance, fundamentally altering how imaging systems were evaluated \cite{barrett2004foundations,barrett2015taskbased}. Quantitative imaging further shifted emphasis from visual appearance to reproducible measurement of physical or biological parameters, requiring explicit attention to definitions, bias, and uncertainty \cite{sullivan2015metrology,kessler2015emerging}. In each case, conceptual innovation helped enable sustained algorithmic progress.

By contrast, early stage medical imaging AI research can often achieve measurable performance gains without revisiting inherited task definitions or evaluation criteria. The availability of benchmark datasets and standardized metrics allows algorithmic optimization to proceed even when conceptual framing is weak or implicit. This capability has accelerated technical development, but it also reduces the pressure to interrogate whether existing problem formulations align with scientific or clinical objectives.

Importantly, this difference reflects structural conditions rather than differences in scientific rigor or intent. Medical imaging AI research operates under distinct data, training, and incentive constraints that make algorithmic advances easier to demonstrate than conceptual ones. Recognizing this historical contrast helps explain why conceptual contributions may appear to be less consistently visible in AI-driven imaging research, despite their central role in shaping long-term scientific progress and clinical impact.

\section{Medical Imaging Versus Computer Vision}

Many methodological norms in contemporary medical imaging AI are inherited from computer vision research developed largely for non-medical applications. While this cross-pollination has enabled rapid technical progress, it also obscures important differences between the two domains that have direct implications for how innovation should be defined and evaluated.

Although there are important exceptions, many computer vision benchmarks involve comparatively stable task definitions, large-scale datasets, and evaluation metrics that align reasonably well with the nominal objective \cite{deng2009imagenet,russakovsky2015imagenet}. In such settings, benchmark improvements can often serve as useful, though still imperfect, proxies for progress, and algorithmic innovation can be assessed with relatively low ambiguity.

Medical imaging AI more often operates under conditions in which labels are noisy or indirect, task definitions are mediated by clinical context, and the relationship between model outputs and meaningful downstream decisions is more difficult to specify \cite{oakdenrayner2020datasets,zech2018variable,roberts2021common}. As a result, success in medical imaging AI cannot be defined solely by visual fidelity or predictive accuracy on retrospective datasets.

These differences place greater weight on conceptual choices in medical imaging research. Decisions about task formulation, ground truth definition, evaluation criteria, and clinical relevance fundamentally shape what algorithmic optimization can achieve. Metrics that are appropriate in generic computer vision may fail to capture clinically meaningful performance, and improvements on standard benchmarks may not translate into improved patient care.

Furthermore, medical imaging AI is subject to constraints that are largely absent in generic computer vision, including safety considerations, regulatory requirements, ethical obligations, and the need for generalization across institutions and populations \cite{fda2026aiml,liu2020consortai,vasey2022decideai}. These constraints amplify the consequences of conceptual misalignment and make explicit reasoning about assumptions and evaluation essential.

Recognizing these domain-specific differences helps explain why conceptual innovation plays a particularly central role in medical imaging AI. While algorithmic advances remain necessary, their scientific and clinical value depends critically on upstream conceptual framing that reflects the realities of medical data, decision-making, and deployment. Approaches that prioritize algorithmic novelty without equivalent attention to these factors risk achieving technical success without corresponding clinical impact. Together, these structural and epistemic differences emphasize the importance of conceptual framing in medical imaging AI and motivate closer examination of how different forms of innovation are recognized and rewarded in practice.

\section{Why Algorithmic Innovation Dominates in Contemporary Medical Imaging AI}

Several observable features of contemporary medical imaging AI research, particularly in method-driven and benchmark-centered contexts, suggest that conceptual contributions are less consistently visible and promoted than algorithmic ones \cite{lipton2019troubling,sculley2018winners}.
These include the structure of prevailing publication formats and review criteria, which prioritize benchmark-driven improvements; the central role of shared datasets and metrics that implicitly fix task definitions; and citation dynamics that favor concrete algorithmic artifacts over conceptual frameworks that diffuse into standard practice \cite{russakovsky2015imagenet,koch2021benchmarks,haldar2026state}. Together, these patterns indicate that algorithmic novelty is often more legible and easier to reward than conceptual advances, even when the latter exert substantial long-term influence.

At the same time, benchmark-driven, algorithm-centric paradigms can translate effectively into clinical value when task definitions, labeling processes, and clinical objectives are tightly coupled. In screening or triage applications, where the goal is to flag studies for expedited human review rather than to produce definitive diagnoses, improvements in conventional metrics such as sensitivity, specificity, or area under the receiver operating characteristic curve can correspond with gains in workflow efficiency and, in some settings, potentially improved patient outcomes \cite{mckinney2020breast,titano2018cranial}. These cases illustrate that benchmark-driven optimization is not inherently flawed; its appropriateness depends on how faithfully the benchmark task reflects the clinical decision being supported.

Training pathways may further reinforce the broader imbalance in how different forms of innovation are emphasized. For researchers whose training is primarily in computer science or engineering, curricula often emphasize optimization, architectures, and implementation more directly than physics-based measurement theory, clinical experimental design, or domain-specific knowledge related to medical applications \cite{nationalacademies2018gradstem}. As a result, problem formulation, assumption analysis, and evaluation design are often treated as fixed quantities rather than as research objects in their own right.

Beyond training pathways, career opportunities outside academia can further incentivize algorithmic innovation. Industry roles in medical imaging and adjacent AI sectors commonly value demonstrated expertise in model development, optimization, and large-scale implementation, as these skills translate directly to deployable systems \cite{stanfordaiindex2025,liang2024academiaindustry}. For early-career researchers navigating uncertain academic prospects, the salience of these opportunities can rationally shape research choices toward algorithmic contributions that are easily legible to non-academic evaluators. This dynamic reflects broader labor-market incentives rather than individual preferences, and it further amplifies the structural advantages of algorithmic novelty over conceptual contributions that may be more difficult to signal outside academic contexts.

Data availability imposes additional constraints on conceptual innovation by shaping which problem formulations are feasible in practice \cite{oakdenrayner2020datasets,clark2013tcia}. Although a growing body of work addresses longitudinal outcomes, treatment decisions, and prognostic modeling, such studies often require substantial institutional access, extended follow-up, and integration of heterogeneous clinical information that are not uniformly available. At the same time, relative to image-domain data, more limited public access to raw measurement data from tomographic imaging modalities---such as projection data, k-space measurements, or list-mode acquisitions---can restrict how inverse problems are studied \cite{knoll2020fastmri,leuschner2021lodopab}. Together, these constraints encourage research questions that align with readily accessible data representations, including image-level reconstructions or standardized benchmarks, even when alternative formulations may better reflect underlying physical processes or clinical objectives.

These dynamics have particular consequences in medical imaging, where task definitions, labels, and evaluation metrics are frequently imperfect proxies for clinical goals. When algorithmic improvement is prioritized without equivalent scrutiny of these upstream choices, progress may appear steady while underlying scientific questions remain unresolved. This emphasis on optimization within fixed framings may help explain why limitations such as poor generalization or limited clinical impact persist despite increasingly sophisticated models.

\section{Conceptual Innovation Carries Disproportionate Risk for Early-Career Researchers}
These same structural features can make concept-driven work especially risky for junior researchers \cite{nationalacademies2018gradstem,lipton2019troubling}. For many early-career researchers, pursuing conceptual innovation can entail greater professional risk than pursuing algorithmic contributions, particularly when review outcomes are uncertain and benchmark-driven improvements provide clearer reward signals. These dynamics reflect commonly observed patterns in contemporary medical imaging AI research rather than universal experiences, and they vary across institutions, subfields, and individual career trajectories.

A central source of this risk is the difficulty of evaluating conceptual contributions consistently \cite{maierhein2024metrics,liu2020consortai,vasey2022decideai}. Algorithmic innovations typically present reviewers with concrete contributions, such as architectures, loss functions, benchmarks, and quantitative improvements, that provide familiar anchors for assessment. Conceptual contributions, by contrast, often involve reframing problems, interrogating assumptions, or proposing alternative evaluation criteria. These elements resist standardized comparison and may be interpreted differently by reviewers with diverse methodological backgrounds, leading to greater variability in review outcomes.

This evaluation ambiguity is compounded by misalignment with prevailing publication norms. Many high-visibility venues in medical imaging AI favor method-centric paper structures that implicitly prioritize technical novelty and performance gains \cite{lipton2019troubling,sculley2018winners,maierhein2024metrics}. Conceptual contributions frequently do not conform to this template, particularly when their primary value lies in clarifying objectives or exposing limitations rather than introducing new computational mechanisms. As a result, conceptually innovative work may be perceived as incomplete or insufficiently technical, even when its scientific implications are substantial.

Conceptual innovation tends to mature more slowly than algorithmic work. Reframing a problem or redefining evaluation criteria often requires sustained discussion, iterative refinement, and the accumulation of evidence across multiple studies, with benefits that may only become apparent as subsequent research adopts and builds upon the new framing. As a result, conceptual advances are frequently most visible in hindsight, after their ideas have been absorbed into standard practice. This temporal mismatch places conceptual innovation at odds with short-term incentives tied to publication frequency, citation metrics, and funding cycles. When reframings become widely adopted, subsequent algorithmic papers may benefit from the conceptual groundwork without explicitly acknowledging its origin, diffusing credit and making conceptual innovation a less reliable source of near-term professional recognition, particularly for researchers who lack established reputations.


Defending conceptual contributions can be particularly challenging due to asymmetries in reputational and institutional leverage within the academic ecosystem. Conceptual innovation often involves questioning established benchmarks, dominant problem formulations, or widely accepted evaluation practices, which may be associated with influential researchers or large research communities. For junior authors, challenging these norms can feel professionally risky, even when critiques are constructive and evidence-based.


Importantly, this risk profile reflects systemic constraints rather than individual reluctance or lack of insight. Many early-career researchers are acutely aware of conceptual limitations in current medical imaging AI practices, but operate within incentive structures that reward safer, more easily legible algorithmic contributions. Recognizing these structural risks is therefore a necessary first step toward identifying how incentives, review practices, and institutional norms might be adjusted to support conceptual innovation without disproportionately disadvantaging researchers at early career stages.

\section{What Is Lost Without Conceptual Innovation}

When conceptual framing is weak, implicit, or insufficiently examined, medical imaging AI research becomes vulnerable to a range of downstream failures that are often misattributed to algorithmic limitations \cite{kelly2019key,roberts2021common}. Poorly specified objectives can lead to misaligned tasks, in which the optimization target does not correspond to the underlying scientific or clinical goal \cite{maierhein2024metrics,lapuschkin2019clever}. In such cases, algorithms may achieve high performance on surrogate metrics while failing to support meaningful decisions, creating a false impression of progress.

Weak conceptual framing also encourages reliance on overconfident evaluation metrics that obscure uncertainty and variability \cite{maierhein2024metrics,liu2020consortai}. Standard benchmarks often compress complex clinical phenomena into single summary statistics, masking heterogeneity across patient populations, acquisition protocols, and disease presentations. As a result, reported performance may overstate real-world reliability, particularly when models are deployed outside the narrow conditions under which they were trained and evaluated.

A further consequence is poor generalization across datasets and institutions \cite{zech2018variable,lapuschkin2019clever,oakdenrayner2020datasets}. Conceptual assumptions embedded in task definitions, label construction, and evaluation protocols often reflect local practices, scanner configurations, or annotation conventions. When these assumptions are not made explicit, models optimized under one framing may fail when transferred to new settings, even if the underlying algorithm is robust. Such failures are frequently interpreted as evidence of insufficient model capacity or training data, when they instead reflect mismatches between conceptual framing and deployment context.

A particularly consequential manifestation of weak conceptual framing arises in  attempts to use deep learning to resolve fundamentally ill-posed inverse problems, such as extreme super-resolution or image reconstruction from highly incomplete measurement data \cite{antun2020instabilities,bhadra2021hallucinations}. In these settings, learning-based methods may produce visually plausible outputs by exploiting statistical regularities in training data, even when the underlying measurement data are insufficient to uniquely and stably determine the imaged object property \cite{antun2020instabilities,bhadra2021hallucinations}. While such approaches can yield impressive performance under conventional fidelity-based metrics, they do not resolve the underlying non-uniqueness of the problem. 
As a result, outputs may reflect learned priors rather than patient-specific information, effectively replacing clinically relevant image content with statistically plausible structures that are not supported by the acquired measurements \cite{bhadra2021hallucinations,haldar2026state}.
Without explicit problem reframing or the incorporation of additional constraints, measurements, or decision-focused objectives that tie model outputs to downstream clinical tasks, clinical trustworthiness may be inherently limited.


These issues directly impede clinical translation \cite{kelly2019key,sendak2020path,vasey2022decideai}. Systems optimized for retrospective benchmarks may not integrate cleanly into clinical workflows, support actionable decisions, or communicate uncertainty in ways that align with clinician reasoning. Algorithmic sophistication cannot compensate for objectives that do not reflect real-world use, and increasing model complexity may exacerbate, rather than resolve, misalignment.

Historically, many of the most durable advances in medical imaging arose from conceptual shifts that redefined what was being measured, how imaging data were interpreted, and how performance was assessed \cite{hounsfield1980computed,lauterbur1973image}. Foundational modalities such as computed tomography and magnetic resonance imaging exemplify this pattern, having emerged from conceptual reframings of image formation and estimation targets that enabled sustained generations of algorithmic development and subsequent conceptual refinements \cite{hounsfield1980computed,lauterbur1973image}.  Task-based image quality assessment, which challenged the assumption that  fidelity alone defines image quality, and the development of quantitative imaging biomarkers, which reframed images as measurable biological signals rather than qualitative pictures, further illustrate how conceptual innovation establishes meaningful objectives and evaluation frameworks that guide and constrain algorithmic progress \cite{barrett2004foundations,barrett2015taskbased,sullivan2015metrology}.


Viewed in this light, some limitations often described as ``model failure'' are better understood as manifestations of upstream conceptual misalignment rather than deficiencies in learning architectures alone \cite{roberts2021common,maierhein2024metrics}.
When models fail to generalize, exhibit unexpected biases, or underperform in clinical settings, the relevant shortcomings often lie in how the problem was framed, which assumptions were made about data and labels, and how success was defined. Recognizing this distinction is essential for directing research effort toward conceptual refinement rather than repeatedly attempting to compensate for misaligned objectives through larger training datasets or increased algorithmic complexity alone.

\section{Illustrative Examples of Conceptual Framing in Medical Imaging AI}

The following examples are not intended as a comprehensive survey of medical imaging AI. Rather, they illustrate three recurring ways in which conceptual framing can shape the meaning of algorithmic progress: by redefining image quality in task-based terms, by redefining intermediate outputs according to their downstream consequences, and by redefining the clinical task itself. In each case, substantial algorithmic innovation can occur within a fixed formulation, but the scientific and clinical significance of that innovation ultimately depends on how the problem is posed and how success is evaluated.

\subsection{Reconstruction and denoising: fidelity metrics versus clinical validity}

In image reconstruction and denoising, algorithmic progress is often measured using appearance-based or pixel-level similarity metrics such as PSNR or SSIM, implicitly treating visual agreement with a reference image as a proxy for clinical value \cite{wang2009mse,wang2004ssim,haldar2026state}. This framing has supported major technical advances, but it also narrows the meaning of success. Diagnostic tasks such as lesion detection, tissue characterization, or decision support do not necessarily track pixel-level similarity, and methods that perform well under conventional fidelity metrics may still suppress diagnostically relevant features or produce outputs whose trustworthiness is difficult to assess \cite{antun2020instabilities,bhadra2021hallucinations,barrett2015taskbased}. A conceptual reframing therefore shifts evaluation away from image appearance alone and toward task-based or decision-preservation criteria, asking not merely whether an image looks correct, but whether it supports the clinical or scientific use for which it is intended.

\subsection{Segmentation: geometric overlap versus downstream consequence}

Automated segmentation is commonly assessed using geometric overlap measures such as the Dice similarity coefficient, which implicitly treats segmentation quality as equivalent to spatial agreement with a reference annotation \cite{dice1945measures,taha2015metrics}. Yet segmentation outputs are often not endpoints in themselves; they are intermediate representations used for treatment planning, quantitative burden estimation, or other downstream decisions. Under this broader view, not all segmentation errors are equally meaningful: some deviations may be clinically negligible, whereas others may materially alter subsequent interpretation or action. The conceptual contribution, then, lies in reframing segmentation quality in terms of downstream consequence rather than geometric agreement alone. This perspective helps explain why benchmark gains in overlap metrics do not always translate into greater clinical utility, and why evaluation strategies must sometimes be redesigned to reflect task-dependent significance rather than uniform pixelwise equivalence \cite{maierhein2024metrics,taha2015metrics}.

\subsection{Diagnosis versus prognosis: label prediction versus unmet clinical need}

Many imaging AI systems are formulated as diagnostic classifiers that predict disease labels from images, and algorithmic innovation in this setting is typically expressed through improved accuracy on curated retrospective datasets \cite{litjens2017survey,roberts2021common}. However, diagnostic prediction is not always the imaging task of greatest unmet value, particularly when diagnoses are already available through established clinical workflows or complementary testing. A more consequential conceptual reframing positions imaging as a predictive and longitudinal measurement tool that supports prognosis, risk stratification, or treatment response assessment \cite{aerts2014decoding,chen2021predicting}. Under this formulation, the central question is no longer whether an algorithm can reproduce an existing label, but whether it can inform future clinical decisions in a way that changes management or improves outcomes. This shift elevates the importance of temporal modeling, uncertainty, and contextual integration, and it illustrates that the most important innovation may lie not in improving performance within an inherited task, but in redefining the task to better match clinical need.

These examples illustrate that conceptual innovation in medical imaging AI often enters not through new computational machinery alone, but through revised definitions of what should be estimated, how success should be measured, and why a task matters clinically.

\section{Rebalancing Incentives: Recommendations for the Medical Imaging AI Community}

Rebalancing conceptual and algorithmic innovation requires coordinated action across the research ecosystem, as no single group controls the incentives that shape research behavior. Meaningful change depends on aligning expectations and reward structures across mentorship, peer review, publication venues, and institutional evaluation. Three changes are especially important: broader review criteria for non-algorithmic contributions, stronger mentorship and institutional protection for concept-driven work, and greater legitimacy for data and validation regimes that enable alternative problem formulations.

Mentors and principal investigators play a central role in shaping how early-career researchers identify impactful problems and perceive professional risk. By explicitly recognizing conceptual contributions as legitimate primary outputs, rather than treating them solely as precursors to algorithmic work, mentors can legitimize problem framing, assumption analysis, and evaluation design as core scientific activities. Senior researchers are also better positioned to absorb risk, for example by sharing responsibility on concept-driven projects or by protecting junior collaborators when conceptual contributions do not yield immediate benchmark gains.

Reviewers play a central role in establishing scientific legitimacy and strongly influence which forms of innovation are recognized. A critical step is distinguishing between the absence of algorithmic novelty and the absence of substantive contribution. Conceptual work can be evaluated by assessing the clarity of problem formulation, soundness of reasoning, generality across methods or modalities, and implications for future research practice. Applying such criteria helps redirect review emphasis away from evaluation practices that are often ill-suited to conceptual contributions, including routine requests for additional ablation studies or incremental algorithmic comparisons that are not aligned with the primary conceptual contribution. Broadening contribution types should not relax evidentiary standards; rather, it should diversify what counts as evidence and how it is justified.

Editors and journals shape research culture through both explicit policies and implicit norms. Dedicated article types, such as Perspectives, Frameworks, or Methodological Analyses, with clearly articulated review criteria can reduce ambiguity for authors and reviewers alike. Greater flexibility in paper structure can create space for synthesis, argumentation, and conceptual clarity that does not fit neatly into method-centric templates, while still maintaining rigorous scientific standards.

Institutions and funding agencies exert longer-term influence through hiring, promotion, and funding decisions. Evaluation criteria that emphasize short-term outputs, publication counts, or narrowly defined technical novelty can inadvertently discourage conceptual work. Recognizing contributions that shape research direction, influence evaluation norms, or enable downstream innovation, often visible only over longer time horizons, would better align incentives with the goals of a mature scientific field. Support for exploratory, interdisciplinary, or theory-driven projects can further reduce the career risk associated with conceptual innovation.

Rebalancing incentives also requires expanding the forms of data and validation that are considered acceptable and valuable. Increased support for curated datasets that include longitudinal outcomes, treatment context, or uncertainty annotations can enable conceptually richer problem formulations, even when such datasets are smaller or more heterogeneous than conventional benchmarks. In parallel, broader availability of raw measurement data from tomographic modalities, through shared repositories, challenge datasets, or controlled-access frameworks, would facilitate exploration of alternative inverse problem formulations beyond image-domain reconstruction.

More generally, broader acceptance of physics-based simulated data, while acknowledging their limitations and appropriately qualifying claims derived from their use, can enable early exploration of new problem framings, task-based objectives, and uncertainty or failure modes that are difficult to study using clinical data alone. It is important to recognize that overly strict reliance on fully realistic clinical data can inadvertently limit conceptual innovation by restricting what can be posed and tested in the early stages of technology development. By legitimizing diverse data representations and validation strategies, the community can reduce the coupling between data convenience and problem definition, enabling conceptual advances that benefit the field of medical imaging.

When conceptual framing, evaluation design, and interpretation are treated as first-class research outputs alongside algorithmic development, the field is better positioned to produce lasting advances that are not only technically impressive, but also scientifically grounded, generalizable, and clinically meaningful.

\section{Conclusion and Outlook}

This Perspective extends prior discussions by making explicit the role of conceptual innovation in shaping task definition, evaluation, and translation in medical imaging AI, and by articulating how these upstream choices interact with algorithmic development and research incentives. Algorithmic innovation has propelled medical imaging AI research forward, but the long-term value of that progress depends on how the field defines its objectives, evaluates success, and recognizes contributions that reshape those foundations. Re-centering conceptual thinking does not diminish rigor; it expands it by ensuring that technical advances are grounded in meaningful problem definitions and evaluation strategies. A mature science of medical imaging AI therefore requires not only better algorithms, but also clearer standards for identifying and rewarding the conceptual advances that make technical progress scientifically grounded and clinically meaningful.

\section*{Acknowledgments}
The author is grateful to colleagues for many helpful discussions that inspired this work.
This work was supported in part by NIH grants P41EB031772, sub-project 6366, and R01EB034249.

\bibliographystyle{IEEEtran}
\bibliography{references}

\end{document}